\documentclass[apj]{emulateapj}
\usepackage{graphicx} 
\usepackage{natbib}
\usepackage{txfonts, color}
\slugcomment{}

\shorttitle{Rb abundances in GCs}
\shortauthors{D'ORAZI ET AL.}
\begin{document} 

\title{Rubidium abundances in the globular clusters NGC 6752, NGC 1904 and 
NGC 104 (47 Tuc)\footnotemark[1]}

\footnotetext[1]{Based on observations taken with ESO telescopes under program 085.D-0205(A)}

%%%
%%%%
%%%%

   \author{Valentina D'Orazi\altaffilmark{2,3}}
   \author{Maria Lugaro\altaffilmark{3}}
   \author{Simon W. Campbell\altaffilmark{3}}
   \author{Angela Bragaglia\altaffilmark{4}}
   \author{Eugenio Carretta\altaffilmark{4}}
   \author{Raffaele G. Gratton\altaffilmark{5}}
   \author{Sara Lucatello\altaffilmark{5}}
   \author{Francesca D'Antona\altaffilmark{6}} 

    \altaffiltext{2}{Department of Physics and Astronomy, Macquarie University, Balaclava Rd, North Ryde, NSW 2109, Australia}
    \altaffiltext{3}{Monash Centre for Astrophysics, Monash University, School of Mathematical Sciences, Building 28, Clayton, VIC 3800, Australia}
    \altaffiltext{4}{INAF Osservatorio Astronomico di Bologna, via Ranzani 1, I-40127 Bologna, Italy}
    \altaffiltext{5}{INAF Osservatorio Astronomico di Padova, vicolo dell'Osservatorio 5, I-35122 Padova, Italy}
     \altaffiltext{6}{INAF Osservatorio Astronomico di Roma, via Frascati 33, I-00040 Monteporzio, Italy}
	   
    \email{valentina.dorazi@mq.edu.au}

\begin{abstract}

Large star-to-star variations of the abundances of proton-capture elements, such as Na
and O, in globular clusters (GCs) are interpreted as the effect of internal pollution 
resulting from the presence of multiple stellar populations. To better constrain this
scenario we investigate the abundance distribution of the heavy element rubidium (Rb) in
NGC 6752, NGC 1904, and NGC 104 (47 Tuc). 
Combining the results from our sample with those in the   
literature, we found that Rb exhibits no star-to-star variations, regardless the
cluster metallicity, with the possible intriguing, though very uncertain, 
exception of the metal-rich bulge cluster NGC 6388. 
If no star-to-star variations will be confirmed for all GCs, it implies that the stellar source of 
the proton-capture
element variations must not have produced significant amounts of Rb. 
This element is observed to be enhanced at extremely high levels in 
intermediate-mass AGB (IM-AGB) stars in the Magellanic Clouds
 (i.e., at a metallicity similar to 47 Tuc and NGC 6388). 
This may present a challenge 
to this popular
candidate polluter, unless the mass range of the observed IM-AGB stars does not
participate in the formation of the second-generation stars in GCs. 
A number of possible solutions are available to resolve this conundrum, also given that
the Magellanic Clouds observations are very uncertain and may need to be revised.
The fast rotating massive stars scenario would not face this potential problem 
as the slow mechanical winds of  
these stars during their main-sequence phase do not 
carry any Rb enhancements; 
however, these candidates 
face even bigger issues such as the production 
of Li and the close over-imposition with core-collapse supernova timescales.
Observations of Sr, Rb, and Zr in metal-rich clusters such as NGC 6388 and NGC 6441 
are sorely needed to clarify the situation.

\end{abstract}

%
%________________________________________________________________

\section{Introduction}

%The striking collection of observational data, both from high-resolution
%spectroscopy and accurate photometric surveys (e.g., \citealt{carretta10b};
%\citealt{piotto12}), acquired in the last years has severely questioned the
%historic notion of globular clusters (GCs) as the best examples of simple
%stellar populations. 

It is now well established that Galactic globular clusters (GCs) host
multiple stellar populations. This is evidenced by large star-to-star
variations in elements affected by proton captures (Na, O, Mg, Al,
hereafter p-capture elements, e.g., \citealt{cohen99}; 
\citealt{ivans99}; \citealt{marino08}; \citealt{carretta10b}), and
in some cases by the splitting of the evolutionary sequences in
colour-magnitude diagrams (e.g., \citealt{anderson09}; \citealt{piotto09}; \citealt{monelli13}; 
\citealt{milone13}). 
The most popular explanation for the existence of 
multiple populations is the internal pollution scenario, which relies on
the occurrence of at least two different episodes of star formation. 
In this picture a portion of the first-generation stars (FG, O/Mg/C-rich and Na/Al/N-poor) underwent nucleosynthesis through the
proton capture (CNO, NeNa, and possibly MgAl) cycles and polluted
the interstellar medium with the processed material via stellar winds. The
second stellar generation (SG, O/Mg/C-poor and Na/Al/N-rich) formed
from this enriched material, probably within about one hundred Myr
(see \citealt{ventura09} and \citealt{gratton12} for a review).

Observational data from high-resolution spectroscopy and accurate
photometric surveys are starting to provide substantial constraints
on the internal pollution scenario. Currently the debate centres around the
fundamental but as yet unanswered question: {\em which of the FG stars were
  the polluters?} The candidates are necessarily the more massive, short-lived FG stars 
that existed early in the GC evolution, although they could not have been supernovae 
because the Ca and Fe-group content is observed to be constant (with a few 
peculiarities such as, e.g., $\omega$ Cen, M22, NGC 1851, see, 
e.g., \citealt{marino09}, \citealt{carretta11}; \citealt{yg08}) and their fast ejecta cannot be 
retained in GCs. The main competitors include intermediate-mass asymptotic giant branch 
stars (IM-AGB; \citealt{ventura01,ventura09}), Super-AGB stars 
(\citealt{ventura11,dercole12}), and fast rotating massive stars (FRMS; 
\citealt{decressin07}), with novae (\citealt{mz12}) and massive binaries 
(\citealt{demink09}) also being suggested.

Apart from the lack of consensus on the stellar source of the pollution, there are also many other open 
issues relating to the chemical abundance differences within GCs. For example, it is presently unclear 
whether the p-capture element distributions show continuous or discrete behaviour (e.g., 
\citealt{marino08}; \citealt{carretta13}), and if these distributions have a possible connection with 
the cluster structural parameters (e.g., mass, metallicity, horizontal branch morphology, see, e.g., 
\citealt{gratton11}; \citealt{bragaglia12}). Due to these ambiguities several observational studies 
have explored other elements, with a view to providing better constraints for the theoretical models. 
This has been done for Li and F (e.g., \citealt{pasquini05}; \citealt{dm10}; \citealt{shen10}; 
\citealt{lind11}; \citealt{mucciarelli11}; \citealt{smith05}; \citealt{yong08c}; \citealt{alves12}; 
\citealt{dorazi13a}; \citealt{delaverny13}) and for trans-iron elements produced by $slow$ neutron 
captures (the $s$-process, e.g., \citealt{armosky94}; \citealt{james04}; \citealt{smith08}). 
These previous works have shown that $s$-process element abundances display a homogeneous abundance 
pattern in the majority of GCs (see also \citealt{dorazi10} who derived the Ba content for a sample of 
more than 1000 stars in 15 Galactic GCs).
Interestingly, the picture is further complicated by the presence of some exceptions: in M22 
\cite{marino09} measured a variation in the $s$-process elements Y, Zr, and Ba correlated with a 
metallicity change, in NGC 1851 Ba has been found to vary from a factor of four to more than one dex 
(\citealt{yg08}; \citealt{villanova10}; \citealt{carretta11}), and in the most spectacular case of $\omega$ Centauri
an increasing trend of Ba and La as a function of [Fe/H] has been revealed by the comprehensive 
high-resolution spectroscopic surveys of \cite{jp10} and \cite{marino11}.

In the current study we focus on rubidium (Rb), an element of mixed $s$-process and $rapid$ neutron 
capture ($r$-process) origin (\citealt{arlandini99}; \citealt{goriely99}). The production of Rb 
during the $s$-process is controlled by the neutron density. If the neutron density reaches higher than 
$\sim 10^{9}$ n/cm$^3$, the $s$-process can proceed through $^{87}$Rb instead of $^{85}$Rb via neutron 
captures on $^{85}$Kr and $^{86}$Rb, two unstable isotopes with half-lives against $\beta$ decay long 
enough to represent ``branching points'' on the path of neutron captures. Because $^{87}$Rb has a magic 
number of neutrons, its neutron-capture cross section is more than an order of magnitude smaller than 
that of $^{85}$Rb (\citealt{heil08}). This results in an enhanced production of Rb with respect to 
neighbouring $s$-process elements, such as Sr and Zr, which are not affected by the operation of 
branching points. Due to this property of the Rb production, the [Rb/Sr] (or [Rb/Zr]) ratio 
spectroscopically observed in AGB stars has been used to derive the neutron density during the $s$-process, 
from which the neutron source and the initial stellar mass can be inferred (\citealt{busso95}; 
\citealt{abia01}; \citealt{garcia06}; \citealt{vanraai12}). AGB stars are well known to be the main 
source of $s$-process elements in the Galaxy. In the standard picture, two neutron sources are present 
in the He-rich shell of AGB stars (\citealt{gallino98}). The $^{13}$C($\alpha$,n)$^{16}$O reaction is 
activated in low-mass ($<$ 4 M$_{\odot}$) AGB stars in radiative conditions and produces low neutron 
densities ($\sim$10$^{8}$ n/cm$^3$), resulting in negative [Rb/Sr] and [Rb/Zr] ratios\footnote{The 
[Rb/Zr] ratio may be positive in the specific case of a neutron exposure so low to not allow the flux to bypass the very first bottleneck on the $s$-process path at the magic nucleus $^{88}$Sr.}. The 
$^{22}$Ne($\alpha$,n)$^{25}$Mg reaction is activated in the convective thermal pulses of IM-AGB stars 
and produces high neutron densities (up to $\sim$10$^{13}$ n/cm$^3$), resulting in positive [Rb/Sr] 
and [Rb/Zr] ratios. These positive ratios are a critical, possibly unique, signature of the $s$-process in 
IM-AGB stars (\citealt{vanraai12}; \citealt{karakas12}).
%, which also imply that in these stars variations can be found in [Rb/Fe] up to 0.5 dex not 
%accompanied by variations in any other $s$-process elements 

Also FRMS are predicted by theoretical models to efficiently activate the $^{22}$Ne($\alpha$,n)$^{25}$Mg reaction 
resulting in the production of $s$-process elements (\citealt{pignatari08}; \citealt{frischknecht12}). 
However, these stars do not experience neutron densities as high as IM-AGB stars, and produce 
negative [Rb/Sr] and [Rb/Zr] ratios\footnotemark[1] (see Figure 1 of \citealt{frischknecht12}).
Furthermore, the $s$-process elements are produced in the late 
evolutionary phases of FRMS and ejected in the interstellar medium during the core-collapse supernova 
explosion. The FRMS scenario of GCs, on the other hand, involves pollution from the winds of 
these stars, which occur previous to 
the explosion. This is required both to allow the material to remain in the GCs, as the supernova 
fast ejecta are not retained by GCs (with the exception of the most massive GCs, e.g., $\omega$ 
Centauri), and to avoid variations in Fe. In summary, the FRMS winds carry variations in proton-capture 
elements but not in the $s$-process elements.

Clearly, observational information on the $s$-process elements, 
and Rb in particular, in GCs is crucial to be compared to theoretical models of the potential 
stellar sources of pollution.
To date Rb abundances have been reported only for a handful of GCs: NGC 6752
(\citealt{yong06}), NGC 6388 (\citealt{wallerstein07}), M4 and M5 (\citealt{yong08a}; \citealt{dorazi13b}),
$\omega$ Centauri (\citealt{smith00}), and NGC 3201 ({\citealt{gw98}). In
  this paper we extend upon the previous investigations and present Rb
  abundances in two more GCs: NGC 1904 and 47 Tuc ([Fe/H]$=-$1.60 dex and
  [Fe/H]$=-$0.72 dex, respectively, \citealt{harris96} -update in
  2010). Furthermore, we report Rb abundances for another 6 giant stars in NGC 6752
  ([Fe/H]$=-$1.54), complementing the previous work by \cite{yong06}. 
The information derived from metal-rich GCs such as NGC~6388 and 47 Tuc is crucial because not only it provides us with Rb abundances for a large range of GC
  metallicities -- covering $\sim$1 dex -- but also it allows us to make a
  direct comparison with recent Rb measurements in IM-AGB stars of similar
  metallicity in the Magellanic Clouds \citep{garcia09}. 

Abundances of other $s$-process elements in the three GCs considered here have been analysed by several 
groups. \cite{francois91} analysed the Ba content for four giant stars in NGC 1904, finding a mean value of 
[Ba/Fe]=0.08$\pm$0.08, while \cite{dorazi10} presented the Ba abundances for a sample of 49 giants, reporting a 
constant value of [Ba/Fe]=0.24$\pm$0.03. NGC 6752 has been previously studied by several authors 
including \cite{james04}, who found 
[Sr/Fe]=0.06$\pm$0.07, [Y/Fe]=$-$0.02$\pm$0.01, [Ba/Fe]=0.18$\pm$0.07 for nine sub-giant and nine turn-off stars, 
and \cite{yong05}, who  
analysed a sample of 38 giants in NGC~6752 and derived mean values of [Y/Fe]=$-$0.02$\pm$0.01, 
[Zr/Fe]=+0.18$\pm$0.02, [Ba/Fe]=$-$0.06$\pm$0.02, [La/Fe]=0.10$\pm$0.01, and [Ce/Fe]=0.27$\pm$0.01. Finally, 
s-process element abundance in the metal-rich GC 47 Tuc were published by\nocite{james04} James et al. (2004, 
[Sr/Fe]=0.32$\pm$0.04, [Y/Fe]= $-$0.03$\pm$0.09, [Ba/Fe]=0.29$\pm$0.07); \citeauthor{alves05} (2005, 
[Ba/Fe]=0.35$\pm$0.05, [La/Fe]=0.05$\pm$0.05); D'Orazi et al. (2010, [Ba/Fe]=0.15$\pm$0.01\nocite{dorazi10}), 
and \citeauthor{worley10} (2010, [Y/Fe]=0.50$\pm$0.04, [Zr/Fe]=0.49$\pm$0.03, [Ba/Fe]=0.34$\pm$0.13, 
[La/Fe]=0.32$\pm$0.03). In summary, all these previous investigations have shown that in these three GCs 
the extremely large variations exhibited in the p-capture elements (Na, O, Mg, Al) are not displayed by any 
$s$-process elements. With the present study we extend these previous investigations 
in relation to the crucial element Rb.

%_____________________________________________________________________________________________________________________________

\section{Stellar sample, observations, and analysis}\label{sec:obs}

We analysed a sample of 15 RGB stars (six in NGC 6752, five in NGC 1904, and four
in 47 Tuc) whose stellar parameters and p-capture element
abundances have been published by Carretta et al. (2007, 2009).\nocite{carretta07}\nocite{carretta09a} 
The clusters under consideration span a large range in 
metallicity ($-$1.5 $\lesssim$ [Fe/H] $\lesssim -$0.7 dex), absolute visual
magnitude (a proxy for the current mass), horizontal branch
morphology, and range/shape of the light-element anti-correlations
(\citealt{carretta10b}). Within each GC we targeted both FG
(O-rich/Na-poor) and SG (O-poor/Na-rich) stars, covering almost the whole
extent of the Na-O anti-correlations (see \citealt{carretta09a}).
	
High-resolution spectra were acquired with
FLAMES-UVES (R=47,000; \citealt{pasquini02}), employing the standard setup
at 860 nm (spectral coverage $\lambda\lambda$6600 \AA$-$10600 \AA),
which includes the Rb~{\sc i} resonance line at
$\lambda$=7800.268 \AA. data 
reduction has been accomplished by the ESO personnel for NGC 1904 and NGC 6752 targets, while we 
performed data reduction for 47 Tuc by means of the ESO FLAMES-UVES pipeline (version 
5.1.0\footnote{Available at \url{http://www.eso.org/sci/software/pipelines/}}), running under {\sc 
gasgano} context, and following the standard procedure (bias subtraction, flat-field correction, 
optimal extraction, and wavelength calibration). The output one-dimensional spectra were then sky 
subtracted, shifted to zero radial velocity and continuum normalised within IRAF\footnote{IRAF 
(Image Reduction and Analysis Facility) is distributed by the National Optical Astronomy 
Observatories, which are operated by the Association of Universities for Research in Astronomy, 
Inc., under cooperative agreement with the National Science Foundation.}. The typical S/N ratios 
are in the range 80-150 per pixel at $\sim$ 7800 \AA.

Rb abundances were determined via spectral synthesis with {\sc MOOG}
(\citealt{sneden73}, 2011 version) and the Kurucz (\citeyear{kurucz93})
grid of model atmospheres, with the overshooting option switched on, 
consistently with \cite{carretta09a}. We note that this choice has negligible impact on our derived Rb abundances. 
The spectral line broadening, assumed
to be Gaussian, was evaluated by fitting the profile of the Ni~{\sc i} line
at $\lambda$=7797.59 \AA. 

We then computed synthetic spectra varying the abundances of Rb until the best match with the observed 
spectrum was achieved. The high excitation Si~{\sc i} feature at $\lambda$=7799 \AA~partially blends the 
Rb line on its left wing. We initially assumed (when available) the [Si/Fe] ratios derived by Carretta 
et al. (\citeyear{carretta07}, \citeyear{carretta09b}) and then optimised those values to better 
reproduce the line under scrutiny here. The agreement between the [Si/Fe] ratios derived by us and 
by Carretta et al. is always within 0.05 dex. By considering the 
typical uncertainty reported by Carretta and collaborators for their Si abundances, we evaluated the impact of 
the [Si/Fe] on Rb abundances being in the range 0.04 - 0.09 dex (depending on the GC and on the star).

Examples of the spectral synthesis are shown in Figure~\ref{f:synth} where we present three stars from our 
sample, each from a different GC. The spectral region around the Rb resonance line at 7800\AA~ is contaminated 
by the presence of several weak CN lines; however, their inclusion has negligible impact on the derived Rb 
abundances, as also discussed in detail by \cite{yong08a} and \cite{dorazi13b}. We 
employed the same linelist as \cite{dorazi13b} who used wavelength shifts and hyperfine structure components 
from \cite{ll76} and assumed the isotopic ratio $^{85}$Rb/$^{87}$Rb~$=3$ (\citealt{tl99}). 
We note that the 
activation of the $s$-process branching points at $^{85}$Kr and $^{86}$Rb 
responsible for enhancing the Rb production in IM-AGB stars is also known 
to produce $^{85}$Rb/$^{87}$Rb ratios less than 3. For [Rb/Fe]~$\sim$1.5 the 
ratio reduces to $\sim 0.3$ (Table 3 of \citealt{vanraai12}). We tested
the effect that of this lower limit has on the spectral synthesis and found that 
the [Rb/Fe] ratio increases by only $\sim+0.2$ dex, which is similar to the measurement uncertainties.
We refer to \cite{dorazi13b} study for further 
details on the atomic oscillator strengths and the abundances for the Sun and Arcturus. In 
Figure~\ref{f:param} we show the [Rb/Fe] ratios for all our sample stars as a function of the stellar parameters 
($T_{\rm eff}$, log~$g$, and microturbulence $\xi$). For NGC~1904, 
due to relatively warmer temperatures of the target stars (see 
Table~\ref{t:results}), we could derive the Rb abundance only for one star (\#98). For the other four stars upper 
limits are given.

\begin{center} 
\begin{figure*} 
\includegraphics[width=14cm]{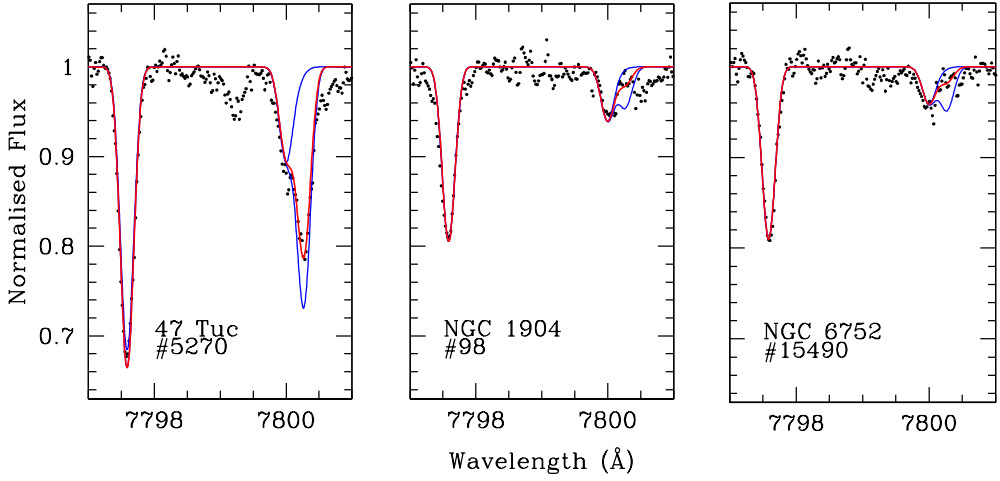} 
\caption{Example of the spectral synthesis for star \#5270 (47 Tuc), 
\#98 (NGC 1904), and \#15499 (NGC 6752). Different syntheses are for A(Rb~{\sc i})=none, best-fit and 
best-fit+0.3 dex values.}
\label{f:synth} 
\end{figure*} 
\end{center} 
\begin{center} 
\begin{figure*} 
\includegraphics[width=14cm]{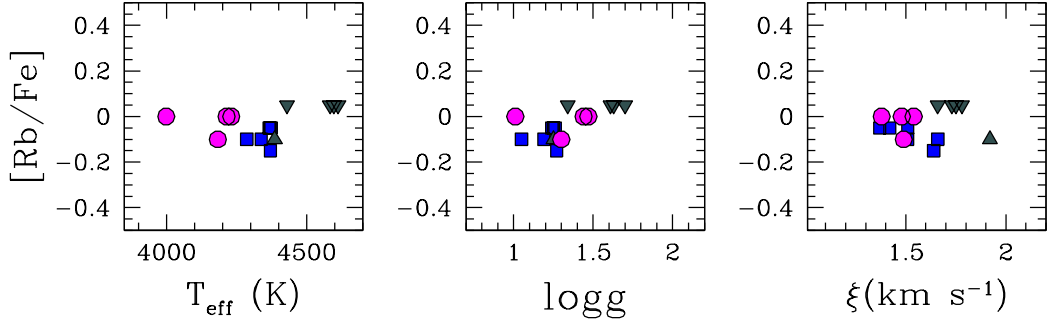} 
\caption{Rb abundances as a function of the effective temperature, 
gravity, and microturbulence for our sample stars. Squares, triangles, and circles are for NGC 6752, NGC 1904, 
and 47 Tuc, respectively. Upside-down triangles represent upper limits.}
\label{f:param} 
\end{figure*} 
\end{center} 
% 
%

%
%\begin{center}
\begin{table*}
\caption{Stellar parameters (from \citealt{carretta07}, \citeyear{carretta09a}) and abundances (Na and O from \citealt{carretta07}, \citeyear{carretta09a}, Si and Rb from the present study).}\label{t:results}
\begin{center}
\begin{tabular}{lcccccccr}
\hline\hline
Star & $T_{\rm eff}$ & log~$g$ & $\xi$       & [Fe/H] & [O/Fe] & [Na/Fe] & [Si/Fe]  & [Rb/Fe] 	\\
     &     (K)       &  (dex)   &(km~s$^{-1}$)&   (dex) & (dex)  &  (dex)  &  (dex)   & (dex)   		\\
\hline
     &               &        &             &        &          &        &          & 		\\
NGC 6752 ~~7627  &   4373 & 1.26 &  1.42  & $-$1.53  & $-$0.08 & ~~~0.72 &  0.45    & $-$0.05        \\
NGC 6752 ~~9756  &   4369 & 1.25 &  1.51  & $-$1.59  & ~~~0.47 & ~~~0.04 &  0.47    & $-$0.05        \\
NGC 6752 15590   &   4366 & 1.24 &  1.37  & $-$1.46  & $-$0.12 & ~~~0.74 &  0.50    & $-$0.05        \\
NGC 6752 23999   &   4286 & 1.05 &  1.51  & $-$1.56  & ~~~0.02 & ~~~0.55 &  0.48    & $-$0.10        \\
NGC 6752 15490   &   4338 & 1.19 &  1.66  & $-$1.55  & ~~~0.42 & ~~~0.13 &  0.40    & $-$0.10        \\
NGC 6752 21828   &   4371 & 1.27 &  1.64  & $-$1.51  & ~~~0.38 & ~~~0.30 &  0.40    & $-$0.15        \\
                 &        &      &        &          &          &   	 &          &        	\\
NGC 1904  113    &   4430 & 1.34 &  1.66  & $-$1.54  & $-$0.50 & ~~~0.55 &  0.30    & $<$0.05      	\\
NGC 1904  185    &   4596 & 1.63 &  1.73  & $-$1.55  & ~~~0.09 & ~~~0.61 &  0.30   & $<$0.05	\\
NGC 1904  181    &   4583 & 1.61 &  1.78  & $-$1.55  & ~~.....  & ~~~0.63&  0.35   & $<$0.05 	\\
NGC 1904  193    &   4612 & 1.70 &  1.75  & $-$1.57  & ~~~0.32 & ~~~0.15 &  0.30   & $<$0.05	\\
NGC 1904  ~~98   &   4386 & 1.25 &  1.92  & $-$1.57  & ~~~0.36 & $-$0.05 &  0.32   & $-$0.10	\\
                 &        &      &        &          &          &   	 &         &         	\\
47 Tuc 13795     &   4183 & 1.30 & 1.49   & $-$0.83  & ~~~0.36 & ~~~0.30 & 0.45    & $-$0.10   	\\ 
47 Tuc 14583     &   4231 & 1.47 & 1.54   & $-$0.68  & ~~~0.34 & ~~~0.43 & 0.40    & ~~~0.00   	\\
47 Tuc 23821     &   4214 & 1.44 & 1.38   & $-$0.78  & ~~~0.39 & ~~~0.44 & 0.37    & ~~~0.00   	\\
47 Tuc ~~5270    &   3999 & 1.01 & 1.48   & $-$0.77  & ~~~0.12 & ~~~0.69 & 0.45    & ~~~0.00   	\\

\hline\hline 
\end{tabular}
\end{center}
\end{table*}
%\end{center}
%
%

Random uncertainties related to best-fit determinations and to the adopted
stellar parameters affect the abundances derived through spectral
synthesis. We found errors in the best-fit procedure (reflecting the S/N of
the spectra and the continuum placement) to be $\sim 0.05$ dex (in [Rb/Fe])
for 47 Tuc and $\sim 0.10$ dex for NGC~6752 and NGC~1904. The
sensitivity of [Rb/Fe] to the input atmospheric parameters was evaluated in the
standard way, i.e., changing a parameter at the time and inspecting the
corresponding variation in the resulting abundance. 
We assumed variations of $\Delta T_{\rm eff}=\pm50$ K,
$\Delta$log~$g=\pm0.2$ dex, $\Delta\xi=\pm 0.1$kms$^{-1}$, and $\Delta$[A/H]=$\pm 0.1$ dex
(Table~\ref{t:sens}); we then computed uncertainties relating to the
stellar parameters adopting errors given by \cite{carretta09a} for NGC 1904
and 47 Tuc (error values as reported in their Table A2 are: 
$T_{\rm eff}$=6K, log$g$=0.04, [A/H]=0.03, $\xi$=0.11 km~s$^{-1}$ for 47 Tuc
and $T_{\rm eff}$=5K, log$g$=0.04, [A/H]=0.03, $\xi$=0.20 km~s$^{-1}$ for NGC 1904)
and by \cite{carretta07} for NGC 6752 (errors are $T_{\rm eff}$=5K, log$g$=0.05, [A/H]=0.05, $\xi$=0.13 km~s$^{-1}$).  The
typical total uncertainties in [Rb/Fe] due to stellar parameters range between 0.03$-$0.05 dex.

\begin{center}
\begin{table}
\caption{[Rb/Fe] dependency on model parameters for two of our stars}\label{t:sens}
\begin{tabular}{lcccr}
\hline\hline
                                &                   &               &           	&	        \\
                                & $T_{\rm eff}$+50  & log~$g$+0.2   & $\xi$+0.1 	& [A/H]+0.1                \\     
\hline
                                &                   &               &            	 &                \\
 $\Delta$${\rm [Rb/Fe]}_{\#5270}$       & 0.06              &$-$0.03        &  $-$0.02  &   $-$0.07       	\\ 
                                &                   &                        &           &		\\
 $\Delta$${\rm [Rb/Fe]}_{\#15490}$      & 0.04              &$-$0.02        & $-$0.02   &   $-$0.08       		\\
 
				&                   &                        &        &                             \\
\hline\hline
\end{tabular}
\end{table}
\end{center}

%
%_________________________________________________________________________________________________________________________________________________
%

\section{Results}\label{sec:results}

In Table~\ref{t:results} we present our resulting Rb (and Si) abundances, along with the stellar 
parameters and Fe, Na, and O abundances by Carretta et al. (2007, 
2009)\nocite{carretta07,carretta09a}.

In NGC~6752 we find the mean Rb abundance to be [Rb/Fe] $=-0.08\pm0.02$
(rms $=0.04$). This compares well with the value of [Rb/Fe] $=-0.17\pm0.06$
reported by \cite{yong06}, implying a difference of $0.09 \pm 0.06$
dex. The small divergence, which is within the measurement
errors (Section~\ref{sec:obs}), is probably due to differences in
temperature and/or metallicity scales: our [Fe/H] ratios are
on average slightly higher and the stars in the sample of \cite{yong06} are cooler
than those analysed here; unfortunately we cannot make 
a direct comparison because none of our sample stars 
is in common with that study.
Based on their findings, \cite{yong06} noted that
the Rb abundances in NGC~6752 appear to be concentrated around two distinct
values, with two stars exhibiting [Rb/Fe] $\simeq -0.02$ and the other
three stars [Rb/Fe] $\simeq -0.25$. However, 
although there was a (possibly bimodal) spread in the Rb data of \cite{yong06}, 
they concluded that it is unlikely NGC~6752 displays a real dispersion
because the uncertainties arising from the weakness of the
Rb~{\sc i} line were quite large. Most importantly, they found no
correlation between [Rb/Fe] and p-capture elements known to vary considerably in GCs. 
Our results deliver further support to this previous work, confirming the lack of an
internal spread in Rb abundances for NGC~6752. 
This abundance pattern is in agreement with other heavy-element abundances, such as 
the second-peak element Ba investigated by \cite{dorazi10} and the third-peak element Pb analysed by 
\cite{yong06}, who concluded that neither elements exhibit any intrinsic spread in this 
cluster. \cite{yong05} determined abundances for a sample of 
38 NGC~6752 giants and found a small variation in the light $s$-process Y and Zr and in 
the heavy $s$-process Ba, positively correlated with the Al variation. However, the detected 
enhancement in the $s$-process elements is at roughly the same level of the measurement uncertainties 
and, if real, might imply a quite peculiar chemical pattern, since according to our data and those of \cite{yong06}, Rb does not display the same kind of internal variation.
It 
could be, however, that the relatively large uncertainties (see also discussion in \citealt{yong06}) 
in the Rb abundances prevent us from identifying a change in its content at the $\sim$0.1 dex 
level.

Because of the relatively warmer temperatures of our stars, only one Rb measurement was possible
for NGC~1904: [Rb/Fe] $=-0.10 \pm 0.12$ dex for star \#98, while upper
limits are provided for the other giants.  The value
obtained for star \#98 is very close to the average Rb abundance in
NGC~6752. These two GCs also share an almost identical metallicity and Ba
abundance: [Ba/Fe] $=0.28\pm0.02$ and [Ba/Fe] $=0.24\pm0.03$ for NGC
6752 and NGC 1904, respectively (\citealt{dorazi10}).

Finally, for the metal-rich GC 47 Tuc we found a mean abundance of [Rb/Fe] $=-0.03\pm0.03$ (rms $=0.05$), i.e., 
a solar [Rb/Fe] ratio and no evidence of internal scatter.\footnote{\cite{milone12} identified 
a multiple sub-giant branch (SGB) in 47 Tuc; the faint SGB might be characterised by different 
$s$-process element abundances (including Rb), however, it accounts only for a very small fraction of the GC 
population and its progeny could not have been targeted here. A dedicated investigation is needed to ascertain 
its heavy-element content; it is worth mentioning in this context that \cite{marino13} detected a significant 
N enhancement in the faint SGB of this cluster.}

\begin{center}
\begin{figure}
\includegraphics[width=9cm]{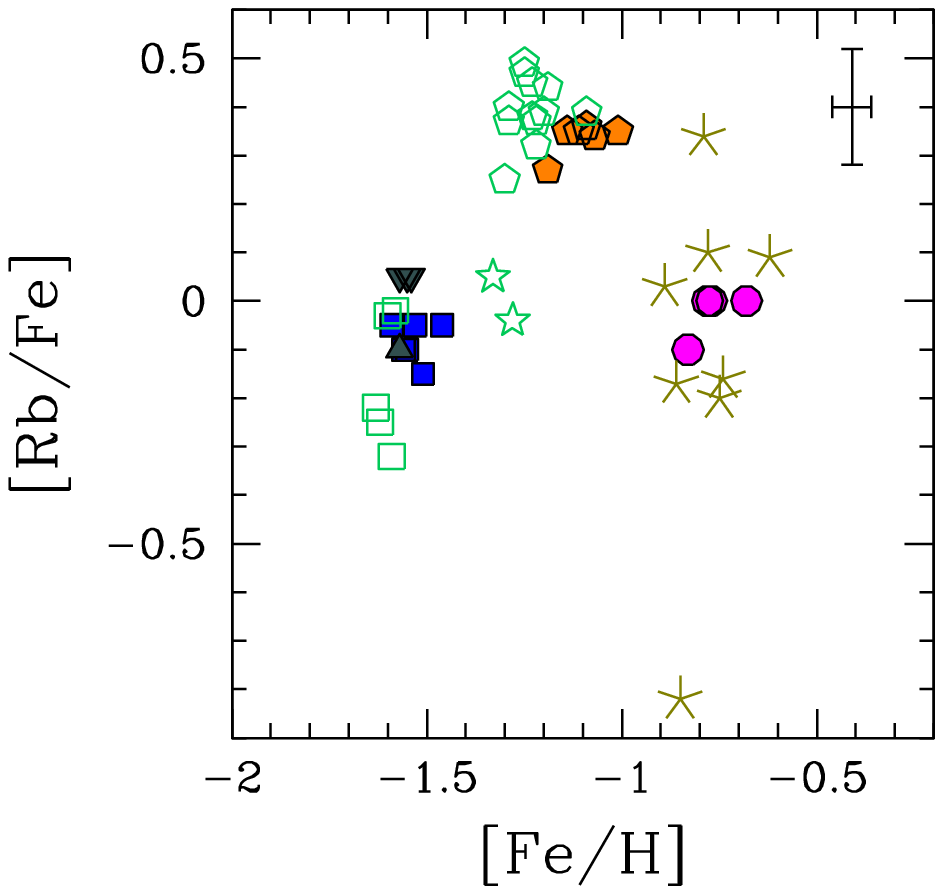}
\caption{[Rb/Fe] versus [Fe/H] for our sample clusters (squares, 
triangles --upside down for upper limits--, and circles are for NGC~6752, NGC~1904, and 47 Tuc respectively) 
along with data for M4 (filled and empty pentagons
  are stars from D'Orazi et al. 2013 and \citealt{yong08a}, respectively),
  M5 (starred symbols; \citealt{yong08a}), NGC 6752 (open squares;
  \citealt{yong06}), and NGC 6388 (asterisks; \citealt{wallerstein07}). 
A representative errorbar is shown in the upper-right corner.}
\label{f:rbfeh}
\end{figure}
\end{center}
%
%
% Rubidium and [Fe/H]
%
%
Figure~\ref{f:rbfeh} displays [Rb/Fe] as function of metallicity for all
our target GCs,
along with values from the literature for NGC~6752, M4, M5, and NGC 6388. We 
deliberately omitted data for $\omega$
Centauri and NGC 3201 because of the non-standard
nature of these systems. In fact, $\omega$ Centauri is known to host
extremely large variations in metallicity and $s$-process elements (\citealt{marino11b}) and 
has been suggested to be the
nucleus of a disrupted dwarf galaxy, while recently \cite{simmerer13}
questioned the mono-metallic nature of NGC 3201.
The first key feature evident in Figure~\ref{f:rbfeh} is that {\em 
  GCs exhibit solar or even slightly sub-solar [Rb/Fe] ratios}, with the
exception of M4. The unusually high intrinsic $s$-process-element content
of M4 has been the subject of many observational studies based on
first-peak and second-peak $s$-process elements (e.g., \citealt{ivans99}; 
\citealt{yong08b}; \citealt{dorazi13b}) and is thought to
reflect a distinct formation and pristine chemical enrichment. The second key feature
in Figure~\ref{f:rbfeh} is that {\em none of the clusters, except for NGC 6388,
  display an internal variation in their Rb content}, within the
uncertainties. 
This constant roughly solar Rb abundance exhibited by GCs represent a crucial result considering 
the large range in metallicity and global structural parameters encompassed. Importantly for the internal 
pollution scenario this also implies that the p-capture element variations detected in all Galactic GCs are 
\textit{not} accompanied by a similar trend in the Rb abundances.  This can be seen clearly in 
Figure~\ref{f:rbnao} where the [Rb/Fe] ratios are plotted as a function of Na and O abundances: internal spreads 
of 0.3$-$0.7 dex (depending on the GC) in Na and of 0.2$-$0.6 dex in O are not accompanied by changes in Rb.

Only in NGC 6388 large Rb variations appear to be present.
\cite{wallerstein07} measured a mean value of 
[Rb/Fe]=$-$0.10$\pm$0.12 (rms=0.34), with one star (\#91) showing an extremely 
low Rb abundance (i.e., [Rb/Fe]=$-$0.82). However, if we do not take into account this star (as 
these authors do in the computation of the mean light $s$-process element abundances), 
the average results in [Rb/Fe]=0.00$\pm$0.07 (rms=0.20). 
\cite{wallerstein07} call the attention on 
two stars in their sample, \#91 and \#361, with very similar atmospheric 
parameters, but almost an oder of magnitude different Rb abundances ([Rb/Fe]=$-$0.82 dex and 
[Rb/Fe]=0.09, respectively). Interestingly, the two stars are also characterised by Na and Al abundances 
correlated with Rb: 
[Na/Fe]=$-$0.07 and [Al/Fe]=$-$0.26 for star \#91 and [Na/Fe]=0.47 and [Al/Fe]=0.53 for star \#361. 
Considering all the data from \cite{wallerstein07}, Figure~\ref{f:rbfeh} shows that 
there is a positive correlation between the Rb and Na 
and Al abundances, however, such a correlation disappears if star \#91 is discarded, and a 
Rb - O anticorrelation is also not present at any statistically significant level. Furthermore, 
the [Rb/H] ratios derived by \cite{wallerstein07} do not exhibit any correlation with their [Zr/H] values. 
Since the Rb uncertainties quoted by \cite{wallerstein07} are $\sim$ 0.3 dex  
(as given by the rms of the mean Rb abundances  
from the two Rb~{\sc i} resonance lines) and their sample stars are characterised by
temperatures notably cooler (down to $T_{\rm eff}$=3500K) 
than those analysed in all the other studies discussed here, 
new measurements of Rb and other $s$-process elements in NGC 6388 are mandatory before any firm conclusion
can be drawn on the presence (or lack) of an internal spread of Rb in this GC.

\section{Discussion}\label{sec:discussion}

Our finding provides a strong constraint on the internal pollution 
scenario for GCs.
The observed lack of a correlation between Rb and
p-capture elements, if confirmed also for NGC 6388,
may be used as an argument against IM-AGB stars as the polluters 
because extremely high Rb 
overabundances (up to 2.6 dex) are observed in IM-AGB stars in the Galaxy 
(\citealt{garcia06}) and even higher values (up to 5 dex) are seen in IM-AGB
stars in the Large and Small Magellanic Clouds (LMC and SMC, respectively, 
\citealt{garcia09}). This is qualitatively in agreement with theoretical 
models, which predict that IM-AGB stars
synthesise and eject substantial amounts of Rb due to the high neutron
density reached during the activation of the
$^{22}$Ne($\alpha$,n)$^{25}$Mg reaction (\citealt{vanraai12}). 

Interestingly, the IM-AGB observational trends of [Rb/Fe] increasing with increasing 
the stellar mass and with decreasing the metallicity are well explained by 
the models (\citealt{vanraai12}) while the observed absolute [Rb/Fe] 
abundances are much higher than predicted. \cite{karakas12} demonstrated that, for 
solar-metallicity stars, [Rb/Fe]$\sim1.4$ could be reached if the final 
stage of mass-loss was delayed resulting in a larger number of thermal 
pulses and increased Rb production. However, a most irksome problem is that 
the observed [Zr/Fe] are roughly solar (within 0.5 dex, \citealt{garcia07}), suggesting no 
production of this element in IM-AGB stars\footnote{In agreement with 
observations of roughly solar Se and Kr abundances (also within roughly 0.5 dex) 
in planetary nebulae of Type II, 
believed to be the progeny of IM-AGB stars (\citealt{sterling08}, 
\citealt{karakas09}).}. This results in [Rb/Zr] ratios much higher than the 
maximum of $\simeq 0.5$ dex allowed by the $s$-process, considering that  
the four elements Rb, Sr, Y, and Zr belong to the first 
$s$-process peak defined by a magic number of neutrons of 50 and are produced 
in similar abundances (within a factor of $\sim$3) for any given 
total time-integrated neutron flux. As discussed in detail by 
\cite{garcia09} and \cite{vanraai12}, there are many difficulties in 
obtaining reliable quantitative abundances in IM-AGB stars due to the 
modelling of the complex, pulsating, dusty atmospheres of these stars. New 
model atmospheres are currently being developed (A. Garc\'ia-Hern\'andez, 
private communication) which may help to shed light on this problem.

In relation to GCs, if Rb production in fact occurs in IM-AGB stars 
at the level seen in the MCs and if the ejecta of these stars contribute to 
the formation of the SG, we should find trace of the presence 
of Rb, particularly in clusters of similar metallicity as the MCs. The
determination of the Rb abundances in 47~Tuc and NGC 6388 is crucial in this respect 
because these clusters have approximately the same metallicity of the SMC ([Fe/H] 
$\approx -0.7$ dex). 

If the lack of a correlation between Rb and 
Na will be confirmed also for NGC 6388, we are in the presence of a conundrum to which possible solutions are:

\begin{itemize}

\item{the mass range of the stars that produce Rb does not play a role in GCs, 
i.e., the IM-AGB polluters could be on average of mass lower or higher (i.e., 
Super-AGBs, as suggested by \citealt{garcia13a}) than the IM-AGB stars observed in the 
MCs;}

\item{the mass-loss rate of IM-AGB stars is fast (\citealt{ventura05b}),
resulting in a small number of
thermal pulses and insignificant $s$-process enhancements (\citealt{dorazi13b}).
Because this solution is the opposite of that proposed by \cite{karakas12} to explain
the [Rb/Fe] observations in IM-AGB stars, it would call for a different evolution of
IM-AGB stars in the cluster environment as compared to the field.}

\item{Alternatively, if updated models of the atmospheres of IM-AGB stars result in a
revision of the [Rb/Fe] ratios in the MCs to much lower values, then
low-metallicity IM-AGB stars must experience a strong mass loss
(see also the recent paper by \citealt{noel13}). In this case they would
avoid
significant production of $s$-process elements.}

\item{Finally, it may be that the huge Rb production observed in the Rb-rich
IM-AGB stars is related to the very final phase of the IM-AGB phase, when
most of the stellar mass has been already lost to the surroundings via
stellar winds. In this case, the observed Rb abundances would be a
signature present only in a small remnant
envelope mass, rather than the overall yield over the whole lifetime of the
star. We note that all the Rb-rich stars observed stars are heavily
obscured, i.e., at the end of their evolution, while massive AGB stars at
the beginning of the AGB phase are super Li-rich and show no Rb enhancement
(\citealt{garcia13b}).}

\end{itemize}

\begin{center}
\begin{figure*}
\includegraphics[width=14cm]{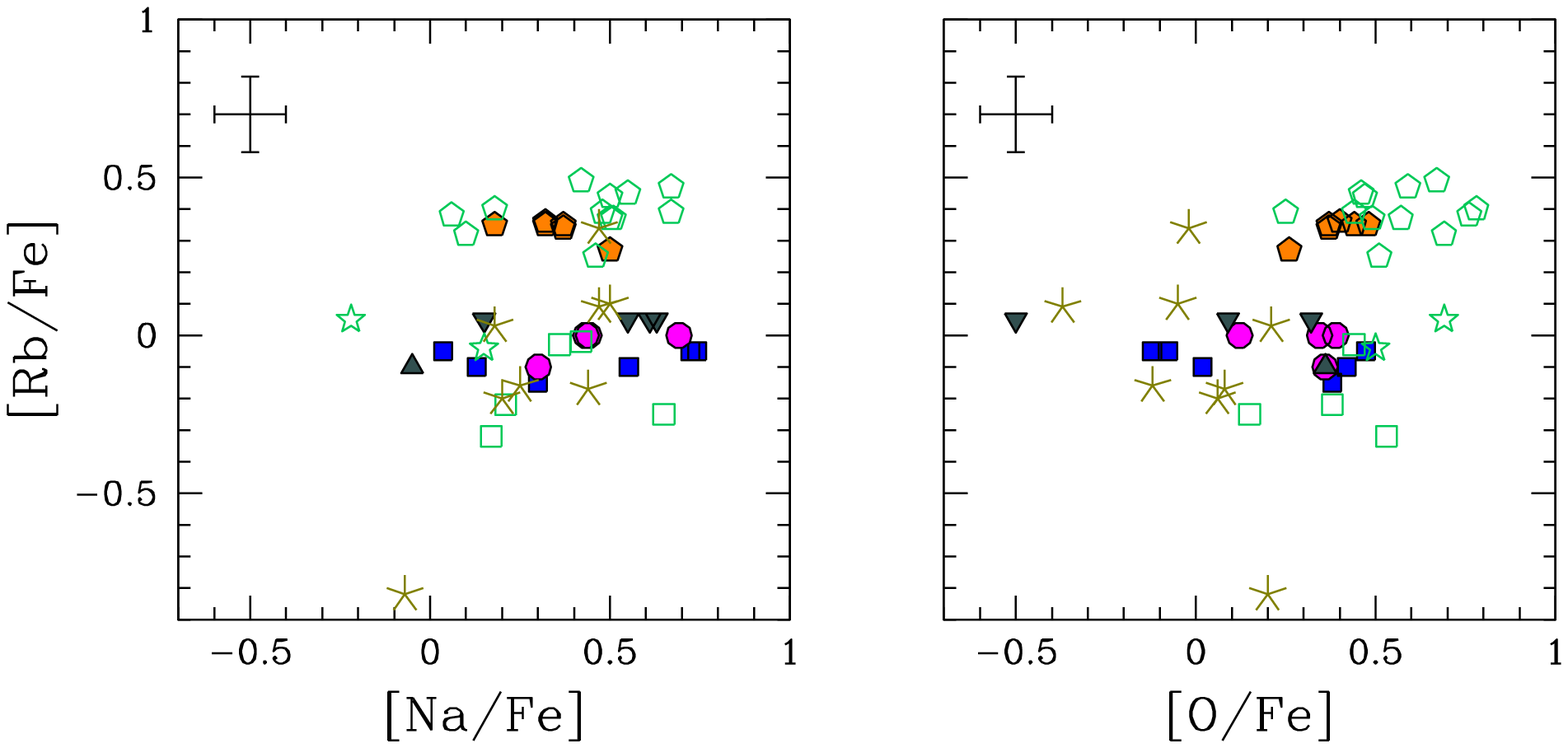}
\caption{Run of Rb with p-capture elements Na and O. Symbols as for Figure~\ref{f:rbfeh}. 
Na and O abundances are from Carretta et al. (\citeyear{carretta07}, \citeyear{carretta09a}), \cite{marino08}, \cite{yong03}, \cite{wallerstein07}.
} \label{f:rbnao}
\end{figure*}
\end{center}
%
%
%
%____________________________________________________________________________________________________________________________________________________
\section{Concluding remarks}\label{sec:conclusions}

Our study on Rb abundances in NGC 1904, NGC 6752, and 47~Tuc suggests that, at odds with p-capture 
elements, this element does not exhibit any internal variation beyond that expected from observational 
uncertainties. This behaviour appears to be true for clusters over a wide range of metallicity, 
however, it urgently needs to be confirmed, or disproved, for the high-metallicity GC NGC 6388. 
If the lack of internal variations is confirmed, 
the internal polluters responsible for the light-element spreads must not have produced 
{\em any} $s$-process elements to a significant extent. Obviously the opposite requirement has to be 
satisfied by the stellar sources responsible for $s$-process variations in ``non-standard'' GCs such as M22 (\citealt{marino09}), NGC 1851 (\citealt{yg08}), $\omega$~Centauri (\citealt{jp10}; 
\citealt{marino11b}; \citealt{dorazi11}), which implies that the type of stars that produced 
the Na-O anticorrelation in all GCs cannot be the same as those that produced the variations in $s$-process 
elements detected in a small sub-sample of GCs (e.g., $\omega$ Cen, M22, NGC~1851).

Comparison with direct observations of Rb in IM-AGB stars in the MCs, i.e., close 
to the metallicity of 47~Tuc and NGC~6388, appears to indicate that IM-AGB stars should produce 
large amounts of Rb, in which case IM-AGB stars might be ruled out as candidates 
for the internal pollution scenario. On the other hand, the FRMS scenario does 
not face this problem as the winds of  
these stars during the main-sequence phase do not carry 
any $s$-process-elements enhancements
(see the Introduction). However, these candidates
face even bigger issues such as destruction/production 
of Li, no depletion in Mg, low fraction of SG stars, and close superimposition 
with core-collapse supernova timescales. 

In conclusion, while the AGB framework has to face major challenges (not only limited to Rb) in 
reproducing the observed features of GCs, we still consider it as the most promising scenario and 
propose a number of possible solutions to solve the possible Rb conundrum in the 
IM-AGB stars scenario, such as e.g.,  a 
different mass range of the AGB polluters, a different IM-AGB mass-loss rate 
(\citealt{ventura05b,dorazi13b}), or the Rb production being limited to the very final thermal pulses.
It should also be kept in mind that the observed [Rb/Zr] ratios are currently orders of magnitude 
higher than values predicted by $s$-process models and might be revised in light of the many 
uncertainties in the model atmospheres of AGB stars (\citealt{garcia09}). 
Finally, we urgently call for new, dedicated observations of 
NGC 6388 to clarify the current picture.
Also, future observations of metal-rich clusters where a large helium difference between the FG and 
the SG is well known (e.g., NGC 6441) will help in putting much stronger constraint on dilution 
and a more accurate estimate of the Rb abundance allowed to be produced by the polluters.

\begin{acknowledgements}

This work made extensive use of SIMBAD, Vizier, and NASA-ADS databases. VD is an ARC Super Science 
Fellow. ML is an ARC Future fellow and Monash fellow.  SWC is supported by an ARC Discovery Project 
grant (DP1095368). We thank the referee for a very careful reading of the manuscript and for 
valuable and helpful comments and suggestions.
 
\end{acknowledgements}

%%\bibliography{draft_rev3}

\clearpage

\end{document}